\documentclass[9pt,twocolumn,twoside]{opticajnl}
\journal{opticajournal} 

\setboolean{shortarticle}{true}

\usepackage{lineno}
\usepackage{upgreek}

\title{On-chip wave chaos for photonic extreme learning}

\author[1]{Matthew R. Wilson}
\author[1]{Jack A. Smith}
\author[1]{Michael J. Strain}
\author[1*]{Xavier Porte}

\affil[1]{Institute of Photonics, University of Strathclyde, Glasgow, G1 1RD, Scotland, United Kingdom}

\affil[*]{javier.porte-parera@strath.ac.uk}

\begin{abstract}
The increase in demand for scalable and energy efficient artificial neural networks has put the focus on novel hardware solutions. 
Integrated photonics offers a compact, parallel and ultra-fast information processing platform, specially suited for extreme learning machine (ELM) architectures. 
Here we experimentally demonstrate a chip-scale photonic ELM based on wave chaos interference in a stadium microcavity. 
By encoding the input information in the wavelength of an external single-frequency tunable laser source, we leverage the high sensitivity to wavelength of injection in such photonic resonators. 
We fabricate the microcavity with direct laser writing of SU-8 polymer on glass. 
A scattering wall surrounding the stadium operates as readout layer, collecting the light associated with the cavity's leaky modes. 
We report uncorrelated and aperiodic behavior in the speckles of the scattering barrier from a high resolution scan of the input wavelength. 
Finally, we characterize the system's performance at classification in four qualitatively different benchmark tasks. 
As we can control the number of output nodes of our ELM by measuring different parts of the scattering barrier, we demonstrate the capability to optimize our photonic ELM's readout size to the performance required for each task. 
\end{abstract}

\setboolean{displaycopyright}{false} 

\begin{document}
\newcommand{\note}{\textcolor{magenta}}

\maketitle



In recent years, there has been a surge in the demand for novel hardware devices to implement neuromorphic computing\cite{Mehonic2024}. 
Photonic integrated circuits (PICs) are a promising platform for implementing hardware-based neural networks because of their small footprint and low power consumption\cite{Farmakidis2024}. 
One type of neural network that are particularly well suited to implement using PICs are feed-forward neural networks (FFNNs)\cite{Shen2017} and in their simplistic version, extreme learning machines (ELMs). 
ELMs can be considered FFNNs with a single hidden layer and random interconnects, where only the readout weights are accessed for training\cite{Huang2006}.

Photonic ELMs have previously been implemented using time-delayed fiber loops\cite{Ortin2015}, scattering media\cite{Saade2016}, multimode fibers\cite{Sunada2020, Tegin2021, Silva2023}, bulk optics\cite{Pierangeli2021}, fiber frequency combs\cite{Lupo2021, Lupo2022, Zajnulina2025}, and photonic integrated circuits\cite{Biasi2023, Rausell2025}. 
The principal mechanism ELMs leverage to solve classification tasks is dimensionality expansion. 
This implies that the hidden layer must be able to expand the input information to a high dimensional parameter space in a non trivial way.
Stadium-shaped microcavities are classical testbed photonic systems for wave chaos\cite{Cao2015}, a physically efficient mechanism for mixing information encoded in light.  
Inside a stadium microcavity, light bounces following chaotic trajectories, which gives rise to high-dimensional mapping of the input information into a complex spatial interference pattern. 
Therefore, such cavities perform physically the hidden layer's task as light propagates and escapes from the cavity.
Those systems have already been studied as photonic reservoir computers (RC) in the past\cite{Laporte2018, UchidaSunadaSciRep2019, Yamaguchi2023}. 
RCs and ELMs share the common learning strategy of optimizing only the output weights, but RCs are based on recurrent neural networks, i.e. they assume a certain degree of fading memory in the network. 
By working in the ELM paradigm, we focus on information processing tasks that do not require memory, such as classification and nonlinear channel equalization. 

\begin{figure}[H]
    \centering
    \includegraphics[width=0.7\linewidth]{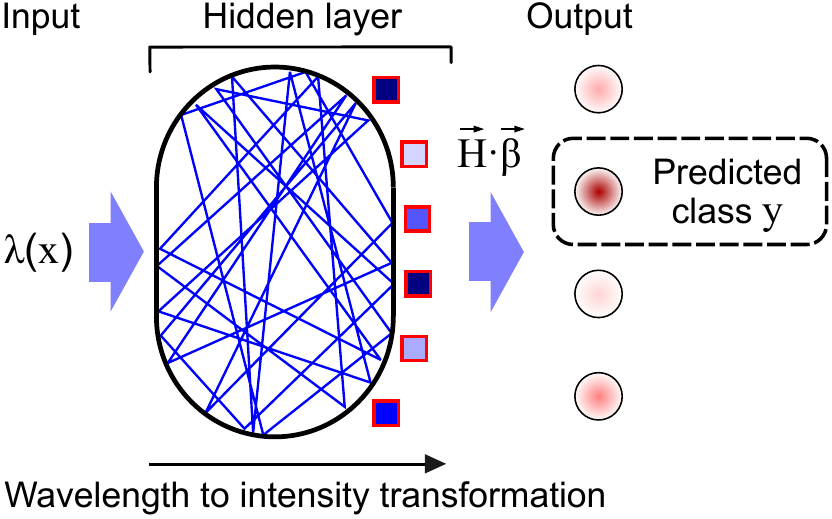}
    \caption{Schematic illustration of our photonic ELM approach. 
    The input information $x$ is encoded in the wavelength of a tunable laser source. 
    The photonic chip transforms the wavelength data $\uplambda(x)$ into a vector representing the hidden layer output nodes $\vec{\mathrm{H}}$. 
    We offline train the readout weights $\vec{\upbeta}$ between the hidden layer and the output layer $y$ for class prediction.}
    \label{fig:intro_fig}
\end{figure}

In this work we fabricate and characterize a polymer stadium microcavity that is surrounded by a barrier designed to scatter the light that laterally escapes from the cavity. 
We then use such system to implement a photonic ELM for benchmark classification tasks of diverse complexity. 
For each task, we characterize the classification accuracy versus the size of the readout layer. 
Our photonic ELM is able to achieve accuracies above $90\%$ in all tasks with a compact chip design, realized with a robust fabrication process and requiring simple optical injection and measurement. 
Figure \ref{fig:intro_fig} illustrates our photonic ELM approach. 
We use the wavelength of a tunable laser source to encode the input data $\uplambda(x)$. 
The on-chip stadium microcavity mixes each wavelength in a complex and unique speckle pattern recovered by an off-chip IR camera. 
We select different positions at the scattering barrier that surrounds the cavity (six in the illustration of Fig. \ref{fig:intro_fig}) and measure the sum of the speckle intensities. 
The number of measured regions corresponds to the size of our ELM's hidden layer $\vec{\mathrm{H}}$ and determines the number of output weights $\vec{\upbeta}$ between that and the output layer of predicted classes. 
Finally, the matrix of speckle intensities together with the vector of target classes $\vec{y}$ are used to train the readout weights using ridge regression. 


\begin{figure}[H]
    \centering
    \includegraphics[width=\linewidth]{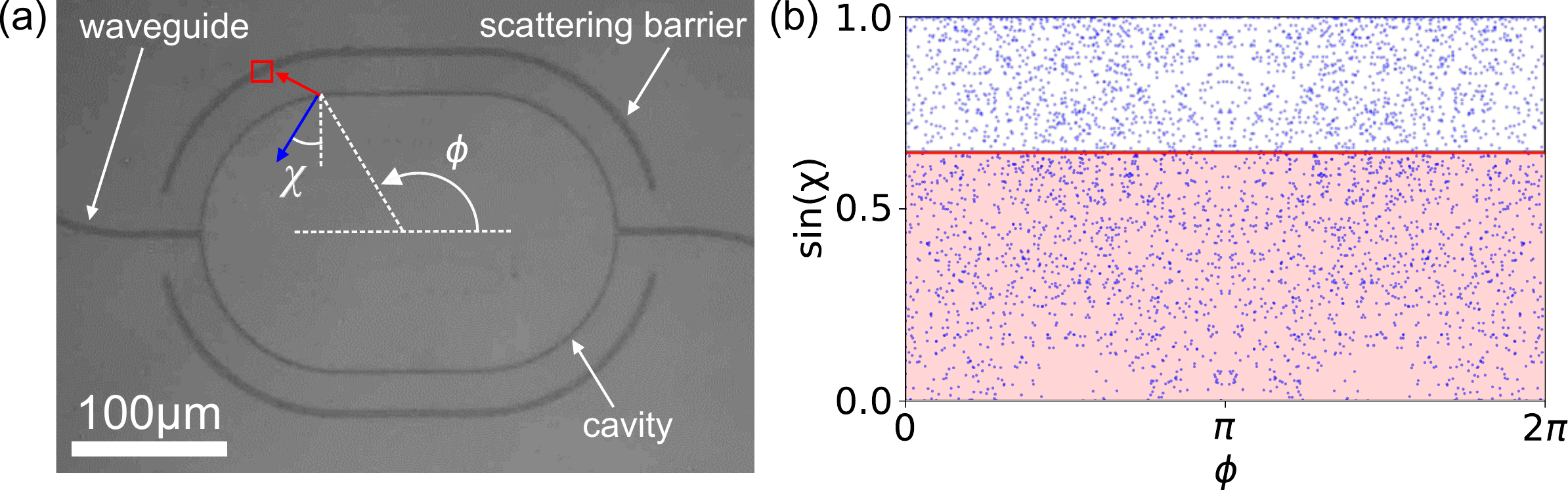}
    \caption{Ray dynamics in stadium billiards. (a) Optical microscope image of the fabricated microcavity with an exemplary illustration of trajectory with reflected (blue) and refracted (red) rays. The red square illustrates a detection node from the surrounding scattering barrier. (b) Poincar\'{e} surface of section for simulated ray trajectories in a stadium chaotic billiard with the same proportions than the microfabricated polymer microcavities. 
    The uniform covering of the phase space is the signature of aperiodic trajectories in the chaotic resonator. 
    The red line denotes the critical angle $\sin(\chi_c)$ for total internal reflection. 
    All angles $\chi < \chi_c$, corresponding to the colored region in panel b, experience total internal reflection.}
    \label{fig:critical_angle}
\end{figure}

The Bunimovich's stadium is a canonical example of non-integrable geometry, which naturally supports nonlinear ray dynamics and wave chaos \cite{DiFalco2012}. 
Such complex dynamics does not require material nonlinearities, making it an inherently energy efficient photonic process for dimensionality expansion. 
We access the internal complex interference field of the cavity by harnessing the light associated with the leaky regions of the phase space. 
Figure \ref{fig:critical_angle}(a) shows an optical microscope picture of the fabricated Bunimovich stadium, a hypothetical ray trajectory leaving the center with an angle $\phi$ will bounce at the boundary with a reflection angle $\chi$. 
If the incidence angle is smaller than the critical angle $\chi_c$, then a part of the light is refracted and scatters onto a $1\upmu\mathrm{m}$-thick barrier placed $30\upmu\mathrm{m}$ away from the cavity boundary. 
Mathematically, this is  when $\sin(\chi) < \sin(\chi_c)$, where $\sin(\chi_c) = \frac{n_{\text{cladding}}}{n_{\text{core}}}$ and $n_{\text{cladding}}$ and $n_{\text{core}}$ are the respective refractive indices of the cladding and core materials.    
In our case, the cladding is air ($n_{\text{cladding}} = 1$), and the core is SU-8 polymer ($n_{\text{core}} \simeq 1.55$).
Figure \ref{fig:critical_angle}(b) depicts the Poincaré surface of section plot for a numerical simulation of a chaotic billiard with the same geometry. 
As can be observed from the evenly distributed points in this phase space representation, our cavity exhaustively explores the phase space of possible trajectories. 
This provides an efficient deterministic mechanism to map input information to a higher-dimensional space, which is a crucial requirement for ELMs performing classification tasks.   
The critical reflection angle $\chi_c$ is indicated with a red line in Fig. \ref{fig:critical_angle}(b), and the leaky region of the phase space (below that angle) is colored in light red. 
Phase space trajectories in the leaky region escape the microcavity and arrive to the outer scattering barrier, where they interfere to create the output of our ELM's hidden layer. 

We follow a straightforward approach based on one-step direct laser writing process to microfabricate the stadium microcavity. 
The structure used in this study was fabricated in SU-8 polymer. 
For this, a glass microscope slide was prepared with a solvent clean and O2 plasma ash. 
A 2$\upmu$m thick layer of SU-8 6002 (Kayaku) was then spin-coated onto the glass slide and soft-baked for 3 minutes. 
The device was created using UV direct laser writing lithography (Heidelberg DLW66+), followed by a further 3 minute soft-bake and development in undiluted PGMEA for 60s. 
The chip was then hard-baked for 30 minutes and the edges cleaved to give side access to the waveguide facets.  
Inspired by a recent paper\cite{Jiang2024} where the authors explore non-Hermitian effects in stadium microcavities, we position a barrier surrounding the cavity to rescatter leaky modes into the vertical direction for imaging with the camera. 
Moreover, injecting from both sides results in a more uniform intensity distribution in the cavity. 


\begin{figure}[ht]
    \centering
    \includegraphics[width=\linewidth]{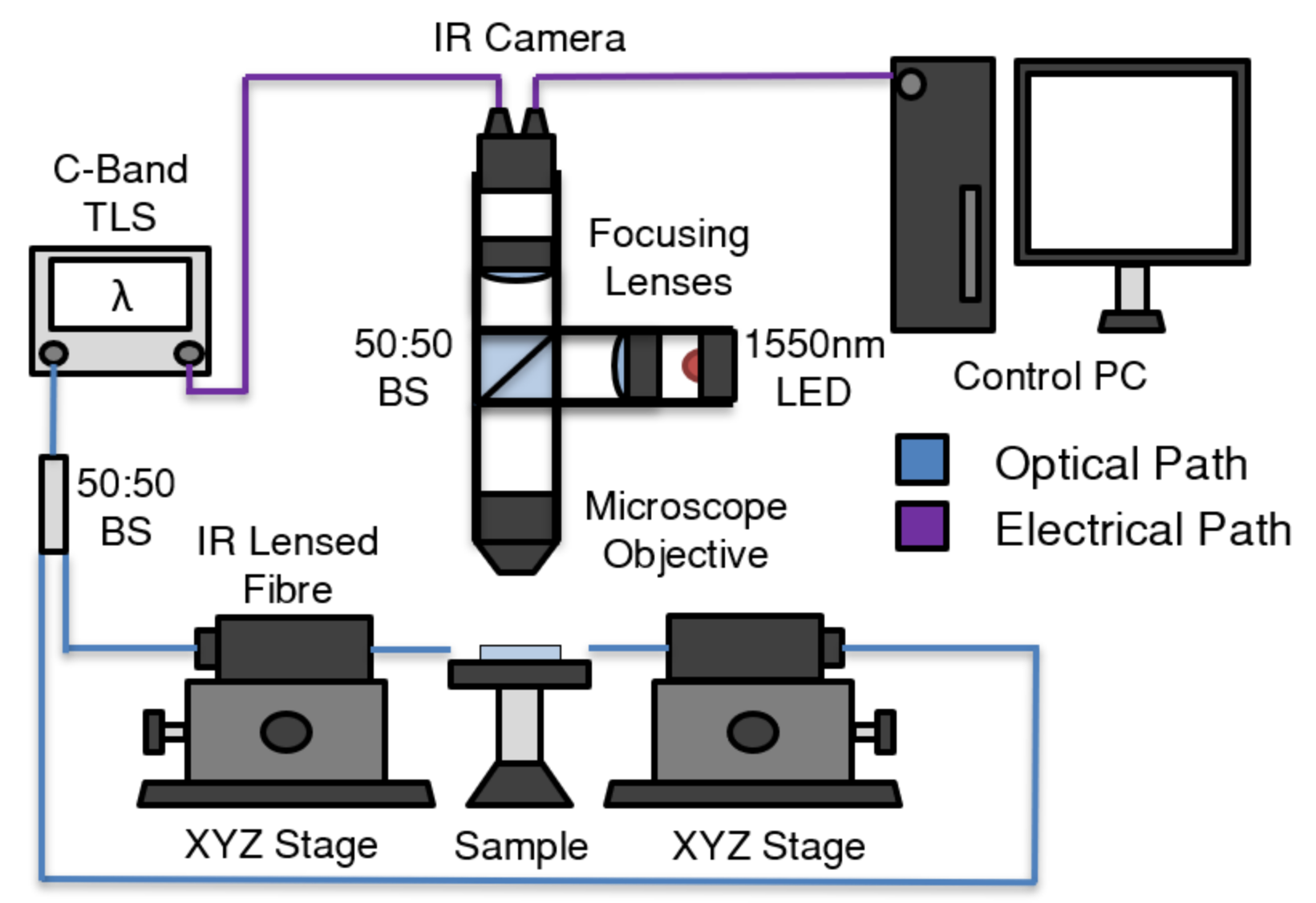}
    \caption{Schematic of the optical characterization setup. Abbreviations stand for TLS: Tunable Laser Source, Bs: Beam Splitter, XYZ: Three-axis micrometric stages, LED: Light-Emitting Diode, IR: Infrared, PC: Personal Computer.}
    \label{fig:setup}
\end{figure}

The optical characterization of the device was carried out using the setup shown in Fig. \ref{fig:setup}. 
The laser source is a fiber-coupled continuous-wave tuneable laser (Santec TSL-570). 
The emitted light is passed through a 50:50 polarization maintaining beam splitter and into two identical single-mode lensed fibers (OZ Optics TPMJ-3A-1550-8). 
The lensed fibers couple light into two single mode SU-8 waveguides on the chip which directly inject the light symmetrically into the cavity. 
The light that is leaked from the cavity acts on the scattering barrier, resulting in a wavelength-dependent speckle. 
Images of the scattering barrier are collected via the imaging setup, which includes a 10X microscope objective (Nikon Plan Fluor 10x/0.30), a 1550nm LED for incoherent illumination of the sample and an IR camera (Lucid Triton SWIR 1.3MP). 

\begin{figure}[H]
        \centering
        \includegraphics[width=\linewidth]{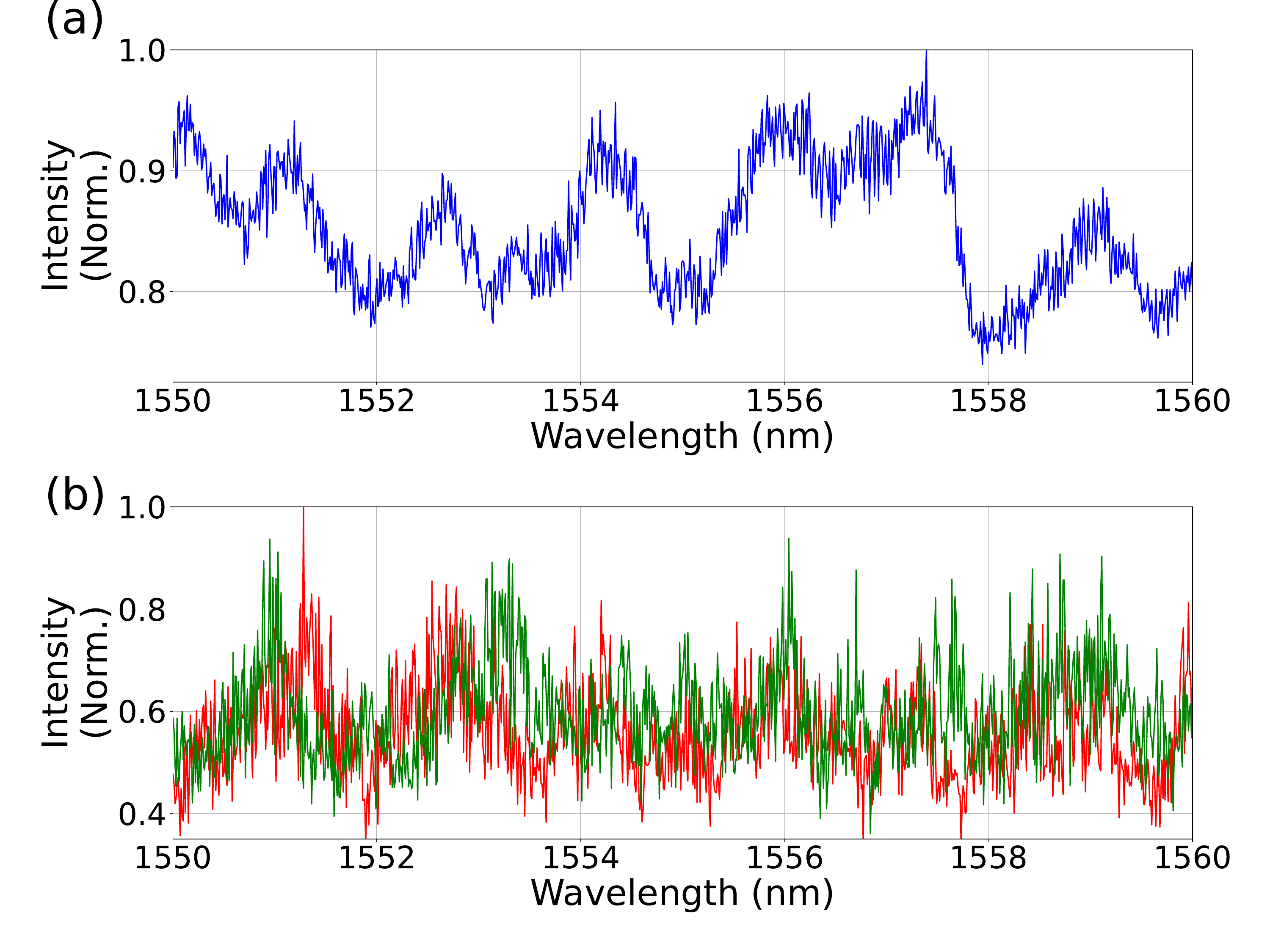}
        \caption{Reconstructed leaky spectrum of the stadium microresonator. 
        (a) Scattered intensity integrated over the whole barrier. 
        (b) Scattered intensity at opposite center positions on the upper and lower parts of the scattering barrier as in Fig. \ref{fig:Cavity_ROIs}.}
        \label{fig:leaky-spectrum}
\end{figure}

To illustrate the sensitivity of the system to small changes in wavelength, we obtained images for each 10pm step in wavelength between 1550nm and 1560nm. 
As mentioned above, we restrict our analysis to the light scattered by the wall surrounding the cavity. 
By integrating the camera pixel values in this region, we can construct a spectra for the leaky light from the cavity. 
The resulting spectra, depicted in Fig. \ref{fig:leaky-spectrum}(a), shows an aperiodic and highly changing structure over the measured 10nm range. 
Crucially, the wavelength dependence strongly changes at different positions around the scattering barrier, as depicted in Fig. \ref{fig:leaky-spectrum}(b). 
The variability of the individual spectra is significantly higher than that of the average and clearly illustrates the space-dependent complex transformation that input information experiences in our system. 

\begin{figure}[H]
    \centering
    \includegraphics[width=\linewidth]{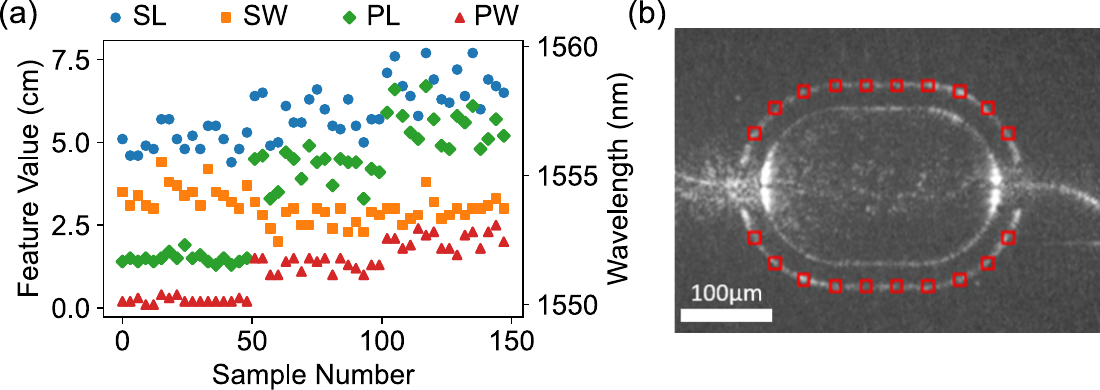}
    \caption{Wavelength to intensity transformation. (a) Plot displaying original feature values of every 3rd sample in the Iris dataset. 
    The length in the original data is linearly mapped onto our tunable laser's wavelength range. The four input features are sepal length (SL), sepal width (SW), petal length (PL), and petal width (PW). 
    (b) Diagram showing how ROIs are defined around the cavity's scattering barrier. Those ROIs are the readout nodes of the ELM's hidden layer. }
    \label{fig:Cavity_ROIs}
\end{figure}

As illustrated in Fig. \ref{fig:intro_fig}, for ELM operation we encode the input information in the wavelength of a tunable laser source and read out the hidden layer's transformation from the intensity scattered at the surrounding barrier. 
The information encoding is done offline in our present demonstration. 
We encode the data samples with a transformation between the sample space and the wavelength space. 
For each classification task, we linearly rescale the components of the m-dimensional input vectors $\vec{x}$ to a wavelength range from 1550nm to 1560nm. 
Figure \ref{fig:Cavity_ROIs}(a) depicts such linear mapping for the particular case of the Iris dataset, with four input features. 
For a given feature $j$, the components of the ELM's response vector $\vec{v}_j(\lambda(x_j))$ is constituted by the integrated intensity inside each of the $n$ ROIs, as indicated in Eq. \ref{eq:info_encoding1}. 
Notably, the readout layer contains the only nonlinearity in our system, as the measured intensity corresponds to the square of the optical field. 

\begin{equation}
\vec{v}_j(\lambda(x_j)) = 
\begin{bmatrix} 
\sum{ROI}_1 \\ 
\sum{ROI}_2 \\ 
\vdots \\ 
\sum{ROI}_{n-1} \\ 
\sum{ROI}_n 
\end{bmatrix} \mbox{ for } j=1,...,m
\label{eq:info_encoding1}
\end{equation}

The feature matrix $\mathbf{H}$ in Eq. \ref{eq:info_encoding2} contains the cavity's response to all $m$ features, containing the transposed vectors of Eq. \ref{eq:info_encoding1}. 
This encoding process is then repeated for each sample $i$ in the dataset $\vec{x}^{ i}$.

\begin{equation}
\mathbf{H} = 
\begin{bmatrix} 
\vec{v}_1(\lambda(x_1))^\mathrm{T} \\ 
\vec{v}_2(\lambda(x_2))^\mathrm{T} \\ 
\vdots \\ 
\vec{v}_{m-1}(\lambda(x_{m-1}))^\mathrm{T} \\ 
\vec{v}_m(\lambda(x_m))^\mathrm{T} 
\end{bmatrix}
\label{eq:info_encoding2}
\end{equation}

In our system, the cavity acts as the hidden layer and the scattering barrier contains its readout nodes. 
We define evenly spaced regions-of-interest (ROIs) on each camera image as readout nodes, as depicted in Fig. \ref{fig:Cavity_ROIs}(b).
The underlying principle of ELMs' operation is to perform a vector-matrix multiplication, solving $\vec{y} = \mathbf{H} \vec{\beta}$. The vector of labels returned by the network is $\vec{y}$, while $\mathbf{H}$ is the output of the hidden layer and $\vec{\beta}$ is the vector of output weights that links the hidden layer and the output classes. 
Training consists in finding the optimal set of output weights such that when an input is presented to the network, the output layer maximally approximates the corresponding labeled target $T$. 
As $\vec{y}$ is simply the linear combination of $\mathbf{H}$ and $\vec{\beta}$, ridge regression is used to find the optimal set of weights. The optimal set of weights is determined by solving $\vec{\beta}=(\textbf{H}^\dagger\textbf{H} + \alpha\textbf{I})^{-1}\textbf{H}^\dagger\vec{T}$ \cite{Biasi2023}, where \textbf{I} is the identity matrix, $\vec{T}$ is the vector of target labels and $\alpha$ is the regularization parameter that helps to prevent underfitting or overfitting the model to the training dataset.

We demonstrate the suitability of our system for ELM computing by performing 4 classification tasks of varying difficulty: Iris flower classification, wine classification, the Wisconsin breast cancer classification task and handwritten digit classification (reduced MNIST). 
Those tasks are directly loaded from the open access library Scikit-Learn \cite{Pedregosa2011}.
The iris flower classification is a classic example of a machine learning benchmark task based on a well known historic dataset. 
The aim of the iris flower task is to classify 150 samples, each described by four features, into one of three species of iris: Setosa, Versicolor and Verginica. 
The four features refer to the length and width of the sepals and petals of the flower. 
The wine classification task is similar to the iris classification task, having slightly higher number of samples and larger feature vectors. 
Here, the aim is to classify 178 samples into 3 different types of wine produced in the same area of Italy. 
Each sample is described by a set of 13 features that are the result of chemical analysis.
The Wisconsin breast cancer classification task is a binary classification task where the two classes are not linearly separable. 
Each sample has a feature vector of size 30, where each element describes a different physical property of the tumor. 
The aim is to classify the 569 samples as benign or malignant.
The handwritten digit classification task is a simplified version of the MNIST digit classification task. 
Each of the 1797 samples are 8x8 pixel images of a handwritten digit from 0-9. 
The integer values of each of the 64 pixels in the image are treated as a feature. 
The aim is to classify the samples by digit.

\begin{figure}[H]
    \centering
    \includegraphics[width=1\linewidth]{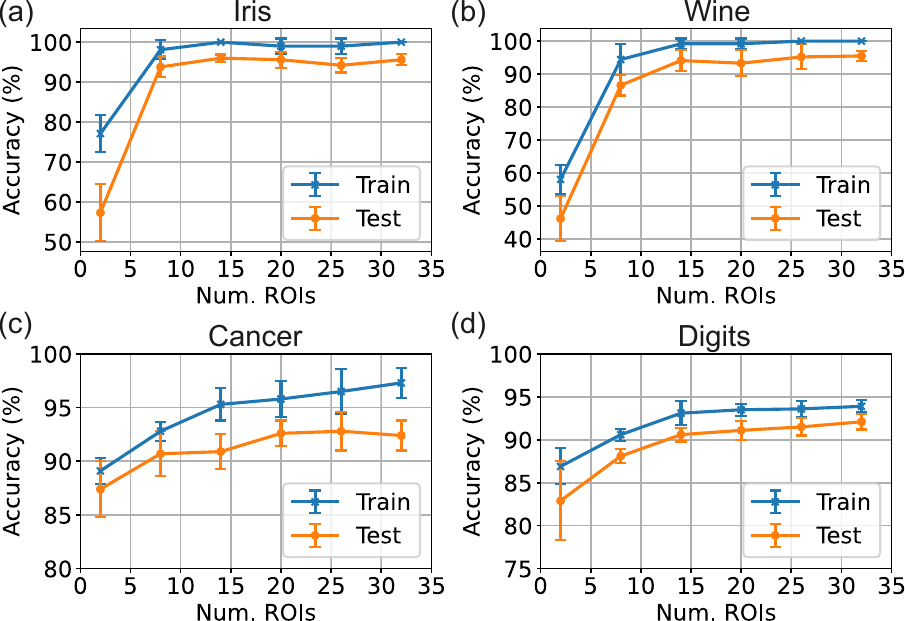}
    \caption{Photonic ELM classification performance for (a) Iris, (b) Wine, (c) Breast cancer, and (d) Hand-written digits tasks.}
    \label{fig:combined-accuracies}
\end{figure}

Figure \ref{fig:combined-accuracies} depicts the performance of our photonic ELM in the above described classification tasks. 
In conventional FFNNs, like ELMs, having control of the size of the readout layer (number of readout nodes) allows the user to control the maximum accuracy achievable by the network and can help to prevent overfitting data\cite{Biasi2023, Rausell2025}. 
With our proposed method, we gain direct control of the number of nodes in the readout layer by changing the number of ROIs used to generate the output vector of the cavity. 
To study the dependence that varying the number of ROIs on the classification accuracy achievable by the model, we perform the tasks above and measure the average accuracies over a 5-fold cross-validation. 
The different datasets were split into training and testing subsets at a ratio of 70:30 and the model was trained by L2 regularization (ridge regression). 
Figure \ref{fig:combined-accuracies} shows the performance results as a function of number of ROIs used in the readout layer for each of the 4 challenged tasks. 
We equidistantly distribute the ROIs along the scattering barrier. 
As we increase the number of ROIs, we observe an increase in average classification accuracy commensurate with increasing the number of readout nodes in our ELM. 
For all tasks in Fig. \ref{fig:combined-accuracies} we see that the classification accuracies saturate above \(90\%\) with 14 ROIs or more. 
Only the binary breast cancer detection task seems to maintain a positive shallow slope in the training accuracy beyond this point. 

In conclusion, we fabricate a polymer stadium microcavity and use it as photonic ELM for wavelength-encoded classification tasks.  
We access the high wavelength sensitivity of stadium microcavities by harnessing the light from their leaky modes. 
Probing the cavity in this way gives control over the number of readout nodes in the ELM.  
Our approach provides a straight forward strategy to scale up the network size to the complexity of the dataset simply scaling up the resonator size and the number of elements accessed from the readout layer. 
We have demonstrated saturation accuracies \(>90\%\) for all tested classification tasks with $<15$ readout nodes. 
Those accuracies are similar to other ELM implementations\cite{Biasi2023,Rausell2025} based on state of the art PIC platforms. 
Notably, recent results demonstrate that the ELM performance can be improved via optimizing the injection data with on-chip active controls\cite{Rausell2025}, therefore a clear direction to increase the accuracies in our system would be to integrate tunable weights with our passive microcavities. 
SU8 has a high thermo-optic tunable coefficient and is therefore a good platform for active control via thermal heaters, potentially overcoming the limitations of this technology reported for silicon PICs leading to significant parameters drift as the overall chip temperature raises because of continuous heat flow\cite{Biasi2022, Biasi2023}.
Further, the saturation accuracies with only $\sim10$ elements in the readout layer is crucial for further hybrid integration of our photonic ELM with chip-scale photodetection modules for fast optoelectronic readout of the network. 
In this respect, a key advantage for our approach is the ready compatibility with grayscale fabrication techniques enabling the microfabrication of off-plane broadband coupling mirrors\cite{Cassells24}.  
Crucially, both polymer-based stadium microcavities and mirrors are wavelength agnostic (our system's material is transparent in a broad range of wavelengths spanning from visible to infrared), making an appealing platform for applications ranging from biophotonics to telecommunications. 

\begin{backmatter}
\bmsection{Funding} 
The Volkswagen Foundation; Fraunhofer UK; Engineering and Physical Sciences Research Council; Royal Academy of Engineering. 


\bmsection{Disclosures} The authors declare no conflicts of interest.

\bmsection{Data availability} Data underlying the results presented in this paper are not publicly available at this time but may be obtained from the authors upon reasonable request.

\bigskip

\end{backmatter}

\bibliography{bibliography.bib}

\bibliographyfullrefs{bibliography.bib}

\end{document}